\documentclass[aps,pre,twocolumn,amsmath,amssymb,10pt]{revtex4-1}

\usepackage{graphicx}
\usepackage{bm}
\usepackage{color}
\usepackage[applemac]{inputenc}

\begin{document}

\title{Collective Dynamics in a Monolayer of Squirmers Confined to a Boundary by Gravity}

\author{Jan-Timm Kuhr}%
 \email{jan-timm.kuhr@tu-berlin.de}%
\author{Felix R\"uhle}%
\author{Holger Stark}%
 \email{holger.stark@tu-berlin.de}%
\affiliation{Institut f\"ur Theoretische Physik, Technische Universit\"at Berlin, Hardenbergstr.~36, 10623 Berlin, Germany}%

\date{\today} 

\begin{abstract}
We present a hydrodynamic study of a monolayer of squirmer model microswimmers 
confined to a boundary by strong gravity using 
the simulation method of multi-particle collision dynamics. 
The squirmers interact with each other via their self-generated hydrodynamic flow fields
and thereby form a variety of fascinating dynamic states 
when density and squirmer type are varied. 
Weak pushers, neutral squirmers, and pullers have an upright orientation. 
With their flow fields they push neighbors away and thereby form a hydrodynamic Wigner fluid at lower densities. 
Furthermore, states of fluctuating chains and trimers, of kissing, and 
at large densities a global cluster exist. 
Finally, pushers at all densities can tilt against the wall normal and 
their in-plane velocities align to show swarming.
It turns into chaotic swarming for strong pushers at high densities. 
We characterize all these states quantitatively.
\end{abstract}

\maketitle

\section{Introduction}
Research of active matter has made much progress in the past decade as documented, 
\textit{e.g.}, by the following review 
articles~\cite{Ramaswamy10, marchetti2013hydrodynamics, Bechinger16, Needleman17, Doostmohammadi18}
but it keeps on challenging natural scientists including physicists. 
One reason is the nonequilibrium nature of active matter and 
the novel and diverse emergent behavior in many of the observed phenomena.
While the microscopic constituents can often be characterized fairly easily, 
a collection of them shows interesting collective dynamics, 
which can be altered by external stimuli or geometrical constraints.
One important example for an active system is a collection of microswimmers. 
They have a propulsion mechanism that enables them to actively move through a fluid 
at low Reynolds number without any applied external force~\cite{lauga2009hydrodynamics}.
However, in the presence of external fields~\cite{stark2016swimming, desai2017modeling}, 
such as flow fields~\cite{sokolov2009reduction, rafai2010effective, zottl2012nonlinear, uppaluri2012flow, zottl2013periodic, clement2016bacterial}, 
light fields~\cite{lozano2016phototaxis, cohen2014emergent}, or 
simple harmonic potentials~\cite{nash2010run, pototsky2012active, hennes2014self}, 
microswimmers show a plethora of fascinating dynamics, 
especially if they are coupled to each other by hydrodynamic interactions~\cite{evans2011orientational, alarcon2013spontaneous, marchetti2013hydrodynamics, saintillan2013active, hennes2014self, yan2015swim, stark2016swimming, zottl2016emergent, 
oyama2016purely, desai2017modeling}.

An ubiquitous example of an external field is gravity, 
which affects every swimmer that is not neutrally buoyant.
Already a range of diverse phenomena has been observed 
such as bound swimmer states~\cite{DrescherGoldstein2009},
polar order in sedimenting swimmers, 
which in turn can be described by an effective 
temperature~\cite{PalacciBocquet2010, EnculescuStark2011, ginot2015nonequilibrium}, 
gravitaxis of asymmetric swimmers~\cite{tenHagenBechinger2014}, 
and inverted sedimentation profiles of bottom-heavy swimmers~\cite{WolffStark2013}.
Very appealing patterns occur during bioconvection~\cite{PedleyKessler1992} and 
more recent work addresses the formation of thin phytoplankton layers in the coastal ocean~\cite{DurhamStocker2009} 
or rafts of active emulsion droplets, 
where the role of phoretic interactions has to be clarified~\cite{KruegerMaass2016, jin2017chemotaxis}.
In realistic settings non-buoyant swimmers will naturally interact with bounding surfaces.
Their impact has already been studied in several works~\cite{llopis2010hydrodynamic, Zottl:2014, jung2014trapping, schaar2015detention, blaschke2016phase, lintuvuori2016hydrodynamic, Ruhle:2018, Thutupalli_2018, shen2018nearwall, shen2019hydrodynamic}.

In recent articles we focussed on single microswimmers in moderate gravitational fields~\cite{Ruhle:2018} 
and on their collective sedimentation~\cite{kuhr2017collective}.
For our studies we used squirmers as model microswimmers~\cite{lighthill1952squirming, blake1971spherical, ishikawa2006hydrodynamic, downton2009simulation}, 
the swimmer type of which can be continuously tuned from pushers over neutral swimmers to pullers.
To fully integrate hydrodynamic interactions between squirmers and 
also between squirmers and bounding walls, 
we employed the method of multi-particle collision dynamics (MPCD), 
a particle-based solver of the Navier-Stokes equations~\cite{gompper2009multi, malevanets1999mesoscopic}. 

In this article we address a monolayer of squirmers that 
forms under strong gravity at the bottom surface of the simulation box
by performing parallelized simulations with up to $10^8$ fluid particles 
and up to several thousand squirmers.
In contrast to our previous work~\cite{Ruhle:2018, kuhr2017collective} where the squirmers can leave the bottom surface
to swim upwards, in this article gravity is so strong that they are constrained to the bottom surface.
Here,
the squirmers either point upwards or 
tilt against the surface normal so that they move along the surface. 
They interact with each other via their self-generated flow fields and 
thereby induce an intriguing variety of dynamic states when density and squirmer type are varied. 
We categorize them in a state diagram.
The most fascinating state is the hydrodynamic Wigner fluid.
It is formed by weak pushers, neutral squirmers, and pullers 
at lower densities due to an effective hydrodynamic repulsion
and it shows a glassy relaxation dynamics.
We also observe states of fluctuating chains and trimers, 
of kissing, and at large densities a global cluster. 
Furthermore, pushers over the whole density range can tilt against the normal and 
the in-plane velocities align to show swarming, 
which turns into chaotic swarming for strong pushers at high densities.

The article is organized as follows.
In Sect.~\ref{sec:methodsandmodel} we present the squirmer model for microswimmers, 
introduce the MPCD simulation method, 
and the relevant parameters of our simulations.
In Sect.~\ref{sec:results} we first present the state diagram for the squirmer monolayer and 
discuss the different states in subsequent subsections.
Finally, in Sect.~\ref{sec:discussion} we summarize our findings and conclude.

\section{Model and Method\label{sec:methodsandmodel}}

In this work we investigate the collective behavior of many squirmers, 
which are coupled to each other and to a confining surface by hydrodynamic interactions. 
In the following we first introduce the squirmer and 
then our hydrodynamic simulation method of multi-particle collision dynamics.

\subsection{The squirmer model swimmer\label{sec:squirmer}}
In our simulations we use the 
squirmer~\cite{lighthill1952squirming, blake1971spherical, ishikawa2006hydrodynamic, 
downton2009simulation, Zoettl18} 
as a versatile model for a microswimmer.
It describes a sphere of radius $R$, which has a prescribed slip velocity field on its surface, 
whereby it propels itself forward without any external force acting on it.
The surface velocity
in the co-moving frame of the squirmer
is given by
\begin{equation}
\label{eq:surface_field}
\mathbf{v}_s(\mathbf{r}_s) = B_1 \left( 1 + \beta \hat{\mathbf{e}} \cdot \hat{\mathbf{r}}_s \right) \left[ \left( \hat{\mathbf{e}} \cdot
\hat{\mathbf{r}}_s \right) \hat{\mathbf{r}}_s - \hat{\mathbf{e}} \right],
\end{equation}
where $\mathbf{r}_s$ is a vector from the center of the squirmer to a point on its surface, 
$\hat{\mathbf{r}}_s = \mathbf{r}_s / R$ is the corresponding unit vector, and the unit vector $\hat{\mathbf{e}}$ 
gives the direction in which the squirmer propels in bulk fluid.
Of course, the surface velocity generates a hydrodynamic flow field
in the fluid in which the squirmer swims.
This flow must satisfy the no-slip boundary condition at bounding walls 
and agree with the surface velocity fields of other squirmers.
Therefore, the squirmer interacts hydrodynamically with other squirmers and bounding walls and thereby experiences
additional linear and rotational velocities.
Equation~\eqref{eq:surface_field} only takes into account the first two terms 
in the expansion of the slip velocity field of the general 
squirmer model~\cite{lighthill1952squirming, blake1971spherical}.
These terms suffice to determine both the bulk swimming speed $v_0 = 2/3 B_1$ and 
the squirmer type through parameter $\beta$ with its characteristic hydrodynamic far field.
Many biological microorganisms, such as \textit{E.~coli} or \textit{Chlamydomonas}, and artificial microswimmers, 
like active droplets~\cite{thutupalli2011swarming, schmitt2013swimming, schmitt2016marangoni, 
schmitt2016active, maass2016swimming, Thutupalli_2018},
can be characterized by these two parameters.
While a squirmer with $\beta = 0$ is a neutral squirmer 
with the hydrodynamic far field of a source dipole ($\sim r^{-3}$), 
$\beta < 0$ and $\beta > 0$ refer to pushers or pullers, respectively, with the far field of a force dipole
($\sim r^{-2}$)~\cite{zottl2016emergent, spagnolie2012hydrodynamics}.
Note that $\beta \ne 0$ also generates a source quadrupole term decaying as $r^{-4}$.

\subsection{Multi-particle collision dynamics\label{sec:mpcd}}
The flow fields generated by squirmers are governed by the Navier-Stokes equations, 
which we solve numerically by employing the particle-based method of multi-particle collision dynamics 
(MPCD)~\cite{malevanets1999mesoscopic, padding2006hydrodynamic, noguchi2007particle, 
kapral2008multiparticle, gompper2009multi},
which includes thermal noise.
In this article we are concerned with microswimmers, which move at low Reynolds numbers. 
In this regime the Navier-Stokes equations reduce to the Stokes equation since inertia is negligible.

In our MPCD simulations the fluid is composed of up to $10^8$ point particles of mass $m_0$,
which perform alternating streaming and collision steps.
In the streaming step the position $\mathbf{r}_i$ of each particle $i$ is updated using its velocity
$\mathbf{v}_i$ and the time step $\Delta t$: $\mathbf{r}_i(t + \Delta t) = \mathbf{r}_i(t) + \mathbf{v}_i\Delta t$. 
In the collision step the simulation volume is divided into cubical cells of edge length $a_0$.
All fluid particles within one cell exchange momentum according to the \mbox{MPC-AT+a} 
rule~\cite{noguchi2007particle}. 
It ensures linear and angular momentum conservation 
as well as a thermalization of particle velocities to temperature $T$.
On average there are $n_\text{fl} = 10$ fluid particles in each collision cell
We choose the duration of the streaming step as $\Delta t = 0.02 a_0\sqrt{m_0/k_B T}$, 
with Boltzmann constant $k_B$, which sets the shear viscosity to 
$\eta = 16.05 \sqrt{m_0 k_B T}/a_0^2$~\cite{noguchi2008transport}.
Note that in a recent work the authors of Ref.~\cite{theers2018clustering} used a larger number of fluid particles per 
unit cell, $n_\text{fl} = 80$, where the MPCD fluid is less compressible. For neutral squirmers confined between to plates, 
they do not observe motility-induced phase separation (MIPS) into a dilute and a cluster phase, which occurs for
$n_\text{fl} = 10$~\cite{Zottl:2014,blaschke2016phase}. Since in our system MIPS does not occur and clustering at medium 
densities is only transient as in the simulations of Ref.~\cite{theers2018clustering}, we used the lower value for $n_\text{fl} = 10$, 
which makes our simulations feasible.

In the streaming step momentum is transferred from the fluid particles to the squirmers.
Fluid particles that enter a squirmer or a bounding surface are repositioned outside of the obstacle 
by updating their positions and velocities.
We apply the ``bounce-back rule''~\cite{padding2005stick} to make sure 
the updated velocity fulfills the surface flow field of eq.~\eqref{eq:surface_field}  
and the no-slip boundary condition at bounding walls, respectively.
Between squirmers and walls as well as between pairs of squirmers also steric interactions are implemented,
which we take into account in a molecular dynamics step.
Further details of our implementation are described in Refs.~\cite{Zottl:2014, blaschke2016phase}. 
Since we simulate large systems with many squirmers, we employ the parallelized version 
of Ref.~\cite{blaschke2016phase}.

As in earlier works~\cite{kuhr2017collective, Ruhle:2018}, we are interested in squirmers moving under gravity
but now make it so strong that all squirmers are constrained to the bottom surface 
with very little variation in $z$-direction as discussed in the beginning of Sect.~\ref{sec:swarming}.
Gravity 
acts on each squirmer with a force $\mathbf{F} = -mg \mathbf{e}_z$. Here $m$ is the squirmer mass and $g$ the acceleration.
Through $g = g_0 (1 - \rho_f/\rho_s)$ it depends on the mismatch of fluid and squirmer densities ($\rho_{f}$ and $\rho_{s}$) 
and the gravitational acceleration $g_0$.
The buoyant squirmer mass then is $m (1 - \rho_f/\rho_s)$. The force $\mathbf{F}$
adds a contribution of $\mathbf{F}/(2m) \Delta t^2 $ 
to the update of the squirmer's position during the streaming step.
Since the influence of gravity on a fluid on the micron scale is negligible,
the update of the fluid particles' positions does not include $\mathbf{F}$.

The MPCD method is known to reproduce analytic results, 
\textit{e.g.}, the flow field around passive colloids~\cite{padding2006hydrodynamic},
fluid friction as a particle approaches a wall~\cite{padding2010translational}, 
the active velocity of squirmers~\cite{downton2009simulation},
or hydrodynamic torques acting on them close to walls~\cite{schaar2015detention}.
Even in systems with many particles, 
such as a dense colloidal suspension under Poiseuille flow, 
it correctly predicts velocity oscillations and particle segregation~\cite{kanehl2017self, kanehl2015hydrodynamic}.

Finally, MPCD resolves flow fields on time and length scales large 
compared to the duration of the streaming step $\Delta t$ 
and the mean free path of the fluid particles, respectively. 
Using a squirmer radius of $R = 4 a_0$, 
we are therefore able to resolve flow fields even if squirmers approach each other closely.

\subsection{Parameters\label{sec:parameters}}
We simulate the collective behavior of up to $N = 3136$ squirmers of radius $R = 4 a_0$.
They are initialized with random orientation and position within a box of height $h = 56 a_0$ 
and a quadratic base with linear extension $L = 112a_0$ or $L = 448a_0$.
We employ periodic boundary conditions in the two horizontal directions ($x$ and $y$) 
while the box is confined by no-slip walls at the top ($z = h$) and bottom ($z = 0$).
Note we expect our box height $h = 14R$ to be sufficiently large so that hydrodynamic interactions with the top boundary 
should be negligible against hydrodynamic interactions between the squirmers moving at the bottom surface. In particular, 
any flow disturbance initiated at the bottom surface decays at least with $1/r^2$. For a mean squirmer distance of $4R$
at a density of $\phi \approx 0.3$ this means that the influence from the top surface is a factor $10$ smaller than the direct 
hydrodynamic interactions between the squirmers.

Similar to our earlier works~\cite{kuhr2017collective, Ruhle:2018} 
there are two velocities that characterize squirmer dynamics. 
These are the bulk sedimentation velocity $v_g = mg / (6\pi \eta R)$ of a single squirmer,
where $m$ is the squirmer mass,
and its swimming velocity $v_0 = 2/3 B_1$.
Their ratio $\alpha = v_0 / v_g$ is a dimensionless number, 
which we fix throughout this work to $\alpha = 0.06$ 
by setting $B_1 = 0.1\sqrt{k_B T/m_0}$ and $mg = 1340 a_0\sqrt{m_0 k_B T}$.
Without activity these squirmers are equivalent to 
passive Brownian particles with the same 
mass $m$, for which we can compute the passive sedimentation length 
$\delta_0 = k_B T / (mg) = 7.5 \cdot 10^{-4} a_0 = 1.9 \cdot 10^{-5}R$.
Hence squirmers experience only very little variability in their vertical position due to thermal fluctuations.
Together with the small $\alpha$ 
this means that squirmers cannot escape from the bottom wall.
The total number of squirmers is sufficiently small so that they form a single 
monolayer with squirmer centers located close to $z = R$. 
We quantify their two-dimensional density by the area fraction $\phi = N \pi R^2 / L^2$.
Note, with monolayer we do not mean a densely packed layer of squirmers but
a collection of squirmers at $z \approx R$ with tunable density.

In our simulations the P\'eclet number $\mathrm{Pe} = v_0 R / D$, 
where $D = k_B T / (6\pi \eta R)$ is the translational diffusion coefficient, 
takes the value $\mathrm{Pe} = 323$.
This means that thermal translational motion is negligible.
Furthermore, we are within the Stokesian regime of hydrodynamics, 
where inertia is irrelevant, as evident from the low Reynolds number of 
$\mathrm{Re} = v_0 R n_\text{fl}/\eta = 0.17$.
Note that real microswimmers move at much smaller Reynolds numbers but the present
value is widely deemed acceptable in particle-based hydrodynamic simulations.

\section{Results\label{sec:results}}
In the following we explore the collective dynamics of a 
monolayer of squirmers confined to a lower boundary by gravity. 
It forms at the beginning of a simulation when the squirmers are rapidly pulled to the bottom wall 
by strong gravity and then equilibrates.
Varying squirmer parameter $\beta$ and area fraction or density $\phi$,
we identify many intriguing states, which are illustrated in Fig.~\ref{fig:parameterstudy}
and by videos S1 to S6 in the supplemental material\dag{}. 
We shortly introduce these states and then discuss them in more detail in the following subsections.
A collection of systems 
at various densities $\phi$ and squirmer parameters $\beta$ is given in video S7.
The upper bound for $\phi$ is given by hexagonally close-packed circles with $\phi_\text{hcp} = \frac{\pi \sqrt3}{6} \approx 0.91$.

\begin{figure}%
\centering%
\includegraphics[width=1\linewidth]{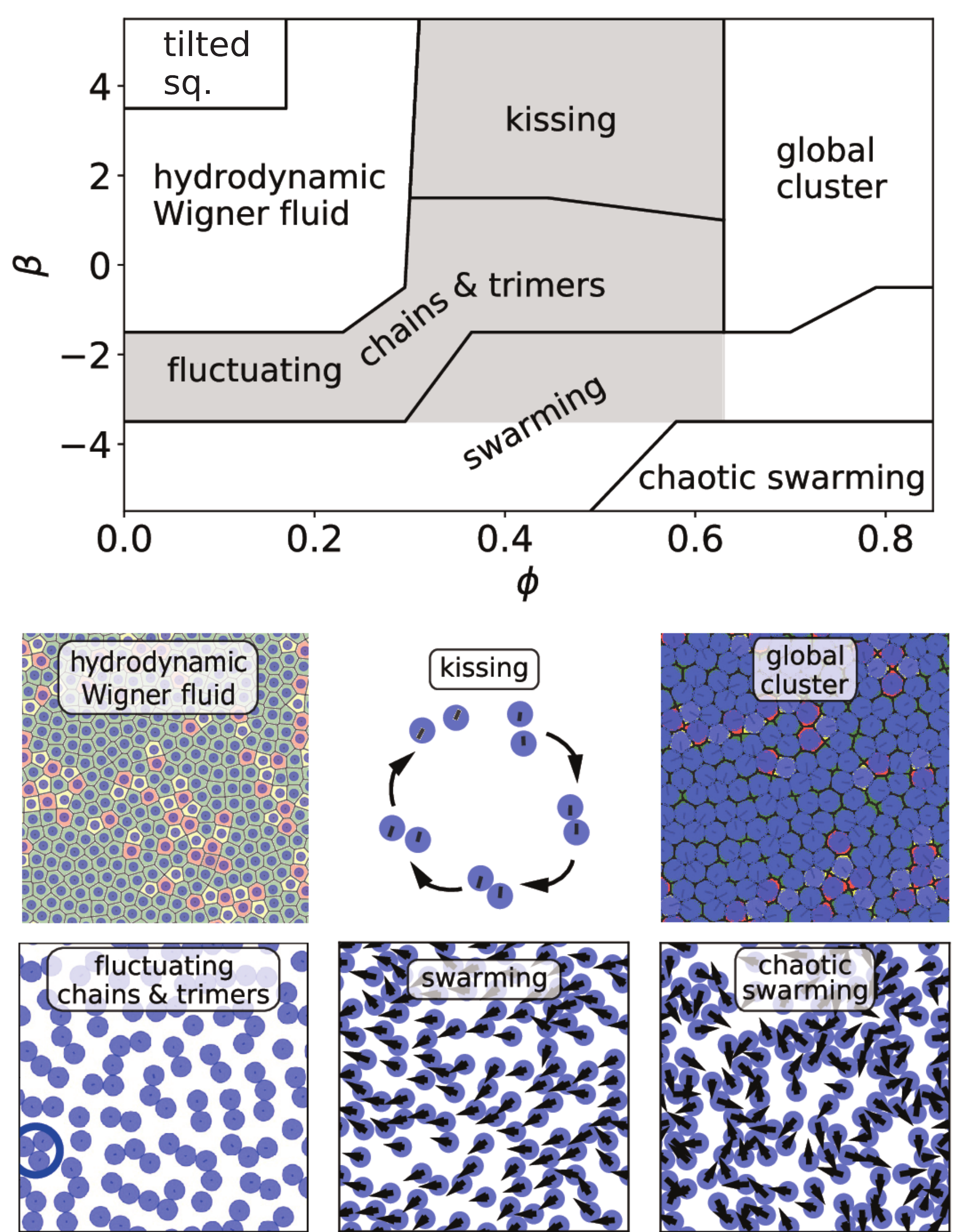}%
\caption{%
\textbf{top:} Schematic state diagram illustrating the collective dynamics of 
a monolayer of squirmers confined to a bounding wall by gravity. 
The different states in the parameter space squirmer type $\beta$ versus area fraction $\phi$ 
are discussed in the main text.
Note, the lines separating the states are schematically drawn.
In the shaded gray area local clusters such as pairs, trimers, and chains are observed.
\textbf{bottom:} 
Snapshots of the observed dynamic states as seen from above.
The \textit{hydrodynamic Wigner fluid} is represented by a Voronoi tessellation, 
where pentagon and heptagon defects are colored yellow and red, respectively. 
The \textit{kissing state} is represented by a single kissing event. 
The red and yellow color in the dense packing of the \textit{cluster state} 
indicate heptagon and pentagon defects. 
The ring in the \textit{fluctuating cluster state} shows a trimer cluster.
In the \textit{(chaotic) swarming state} the arrows point along the orientation of the squirmers.
The supplemental material\dag{} provides videos S1 to S6 of all the 
relevant
states and video S7, which shows a collection of the dynamic states in the whole state diagram.
}%
\label{fig:parameterstudy}%
\end{figure}

For all densities pushers ($\beta < 0$) exhibit swarming (Sect.~\ref{sec:swarming}).
Their swimming direction tilts against the vertical so that they move in the horizontal plane.
While strong pusher at low densities and weaker pushers at large densities swarm with a common direction, 
strong pushers at large densities show chaotic swarming. 
Adjacent in the state diagram is a region at small and medium densities,
where neutral squirmers and weak pushers/pullers form 
pairs, trimers, larger clusters, and chains, 
which form and break up due to stochastic fluctuations (Sect.~\ref{sec:clusters}).
In contrast, stronger squirmer pullers at medium densities 
perform a deterministic maneuver which we term ``kissing'' (Sect.~\ref{sec:clusters}).
They approach each other, form pairs or trimers, 
then orient away from each other and thereby separate again.
At large densities fluctuating and kissing clusters enter a state, 
where one global cluster forms in which squirmers hardly move or behave rather dynamic with increasing $\beta$.
However, the most intriguing state is the hydrodynamic Wigner fluid, 
which weak pushers, pullers, and neutral squirmers form at low to medium densities
(Sect.~\ref{sec:hexorder}).
Due to hydrodynamic repulsion they form local hexagonal order 
while long-range translational and orientational order does not occur. 
We start with describing this state in more detail 
in Sect.~\ref{sec:hexorder}.

\begin{figure}%
\centering%
\includegraphics[width=1.0\linewidth]{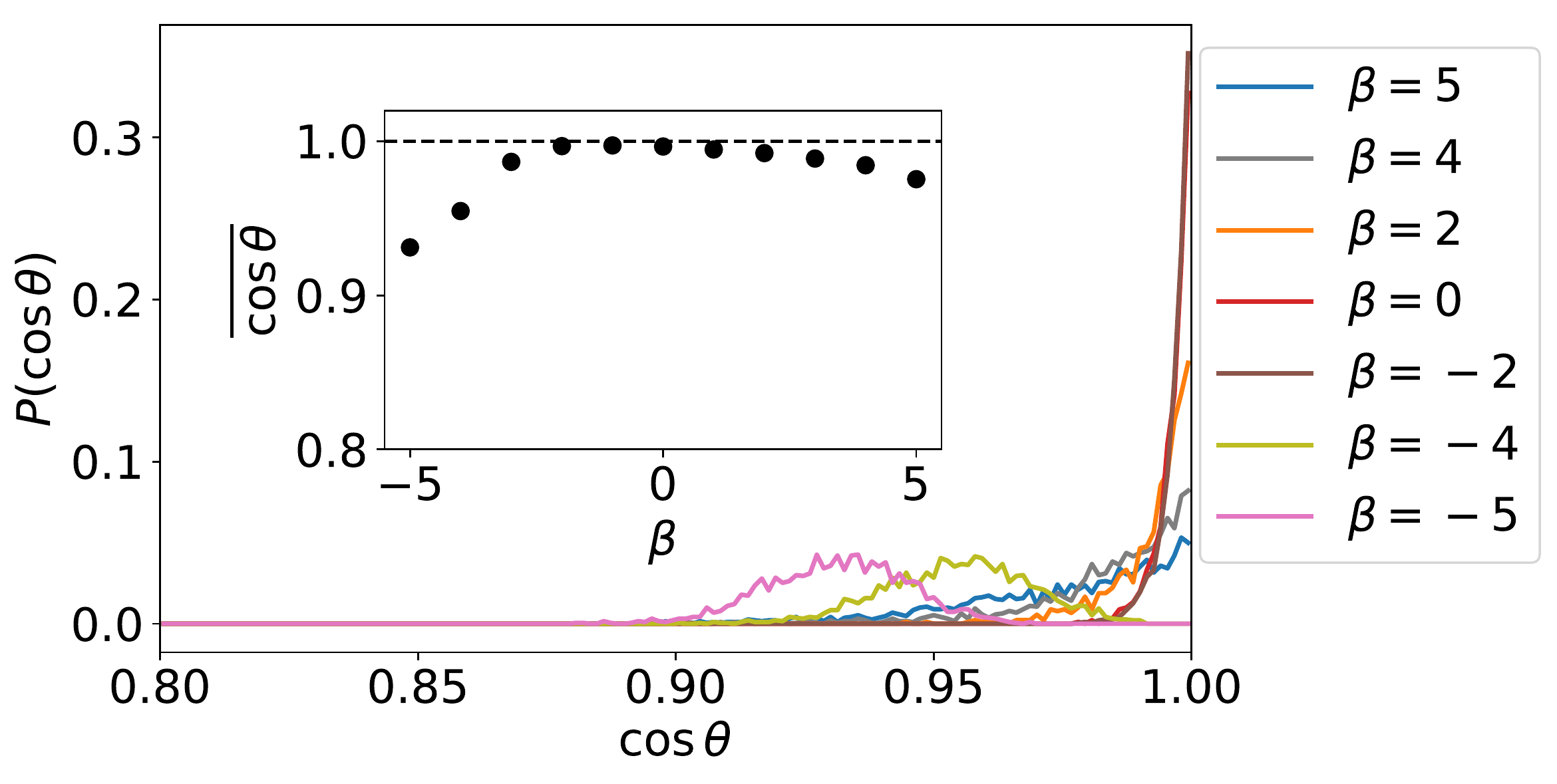}%
\caption{%
Probability
distributions for the orientation of single squirmers confined to a wall by gravity
for different $\beta$.
$\theta$ is the angle between the surface normal and the squirmer orientation $\hat{\mathbf{e}}$.
The inset shows the mean orientation for different $\beta$.
}%
\label{fig:single_sq_orientation}%
\end{figure}

A first understanding of the observed phenomenology at low densities is provided by the probability
distributions for the orientation of single squirmers and their mean orientation plotted in 
Fig.~\ref{fig:single_sq_orientation}. 
Strong pushers ($\beta = -4,-5$) have a maximum of the orientational
distribution at $\theta \ne 0$, where $\theta$ is the angle between the wall normal and the squirmer's orientation. 
Since they are tilted against the wall normal, they move in the horizontal plane and their collective motion initiates swarming.
The tilted orientation is expected for pushers at heights between the validity of lubrication theory and far-field 
approximation~\cite{lintuvuori2016hydrodynamic, Ruhle:2018}. As we will discuss in the beginning of Sect.~\ref{sec:swarming}, 
strong pushers are not sitting directly on the surface. For all other $\beta$ single squirmers are oriented on average along 
the wall normal in agreement with the fluctuating cluster state and the hydrodynamic Wigner fluid. The orientations of 
strong pullers ($\beta = 4, 5$), however, strongly fluctuate about the wall normal and instead of the Wigner fluid they form a state
``tilted squirmers'', which we shortly introduce in the paragraph before Sect.~\ref{subsec.structural}.

\subsection{Hydrodynamic Wigner fluid \label{sec:hexorder}}
\begin{figure}%
\centering%
\includegraphics[width=0.7\linewidth]{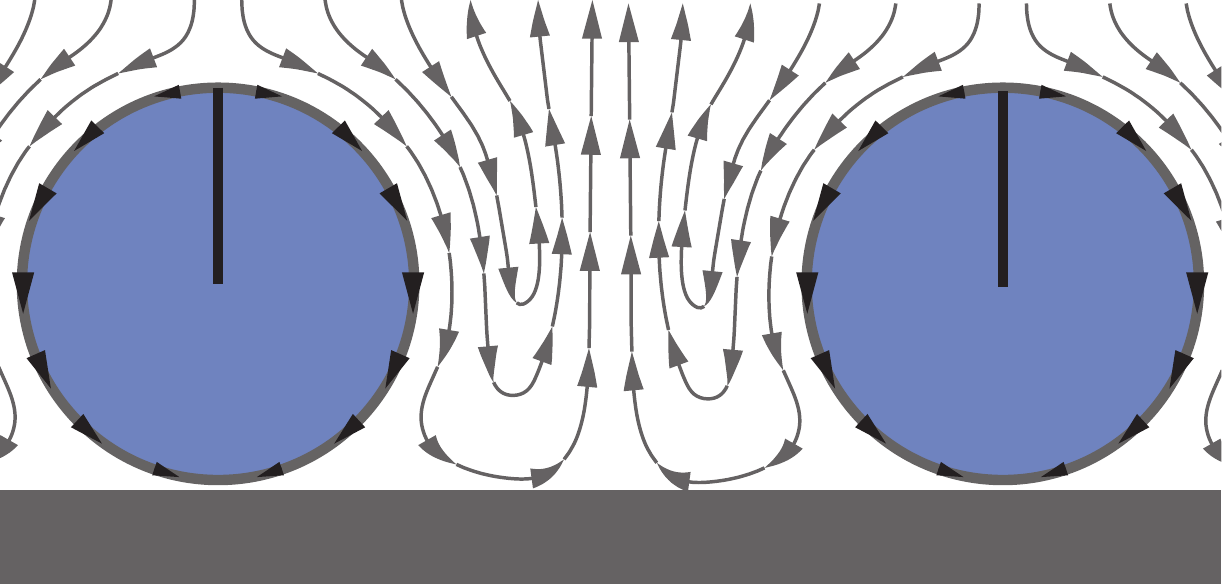}%
\caption{%
Schematic flow profiles between two neutral squirmers repelling each other.
}%
\label{fig:repulsion_scematic}%
\end{figure}%

The state of the hydrodynamic Wigner fluid is most clearly demonstrated for $\beta = 0$ and $\phi = 0.26$ 
by video S8 in the supplemental material\dag{}.
Here, neutral squirmers are well separated from each other and local hexagonal order is visible. 
They fluctuate about their mean position as in a colloidal crystal (see, \textit{e.g.}, Ref.~\cite{zahn1999two}). 
The squirmers have an upright orientation 
and push nearby fluid downwards, which due to the bounding surface pushes nearby squirmers away (see 
Fig.~\ref{fig:repulsion_scematic}). 
Due to this long-range hydrodynamic repulsion local hexagonal order forms.
Obviously, the repulsion is enhanced for pullers, which pull additional fluid downwards and to the side. In contrast, 
pushers at their back draw fluid in and thereby destabilize the Wigner fluid so that with decreasing $\beta$ a transition
to the state of fluctuating chains and trimers occurs (see Fig.~\ref{fig:parameterstudy}, top).

When thermal fluctuations tilt the swimming direction, a squirmer moves towards its neighbors until their flow fields
and the hydrodynamic interaction with the wall rotate the orientation back to normal. 
This causes the clearly visible fluctuations in video S8. Assuming perfect hexagonal order, the squirmer distance 
in units of the squirmer radius is approximated as 
\begin{align}
\frac{d_\text{hex}(\phi)}{R} = \!\sqrt{\frac{2\pi} {3^{1/2} \phi}} \, ,
\end{align}
which for $\phi = 0.26$ amounts to $d_\text{hex}/R \approx 3.75$. Here, the local ordering is most pronounced. 
At low densities hydrodynamic repulsion and thus hexagonal ordering is weaker.
In the other direction when density increases towards $\phi = 0.3$, 
approaching squirmers start to touch each other 
and the system enters the states of fluctuating chains/trimers and kissing (see Sect.~\ref{sec:clusters}).
This implies that short-range interactions between the squirmers are effectively attractive.  
An exception are pullers with $\beta = 4, 5$, which are tilted against the normal so that they move backwards
(note the weak maximum at $\cos \theta = 1$ of the orientational distribution function for a single squirmer
in Fig.~\ref{fig:single_sq_orientation}).
At low $\phi < 0.2$ (see state tilted squirmers in the state diagram of Fig.~\ref{fig:parameterstudy}) this leads to 
a long-range hydrodynamic attraction and a short-range repulsion, where local hexagonal ordering cannot form.
We will not discuss this state further.

\begin{figure}%
\centering%
\includegraphics[width=0.9\linewidth]{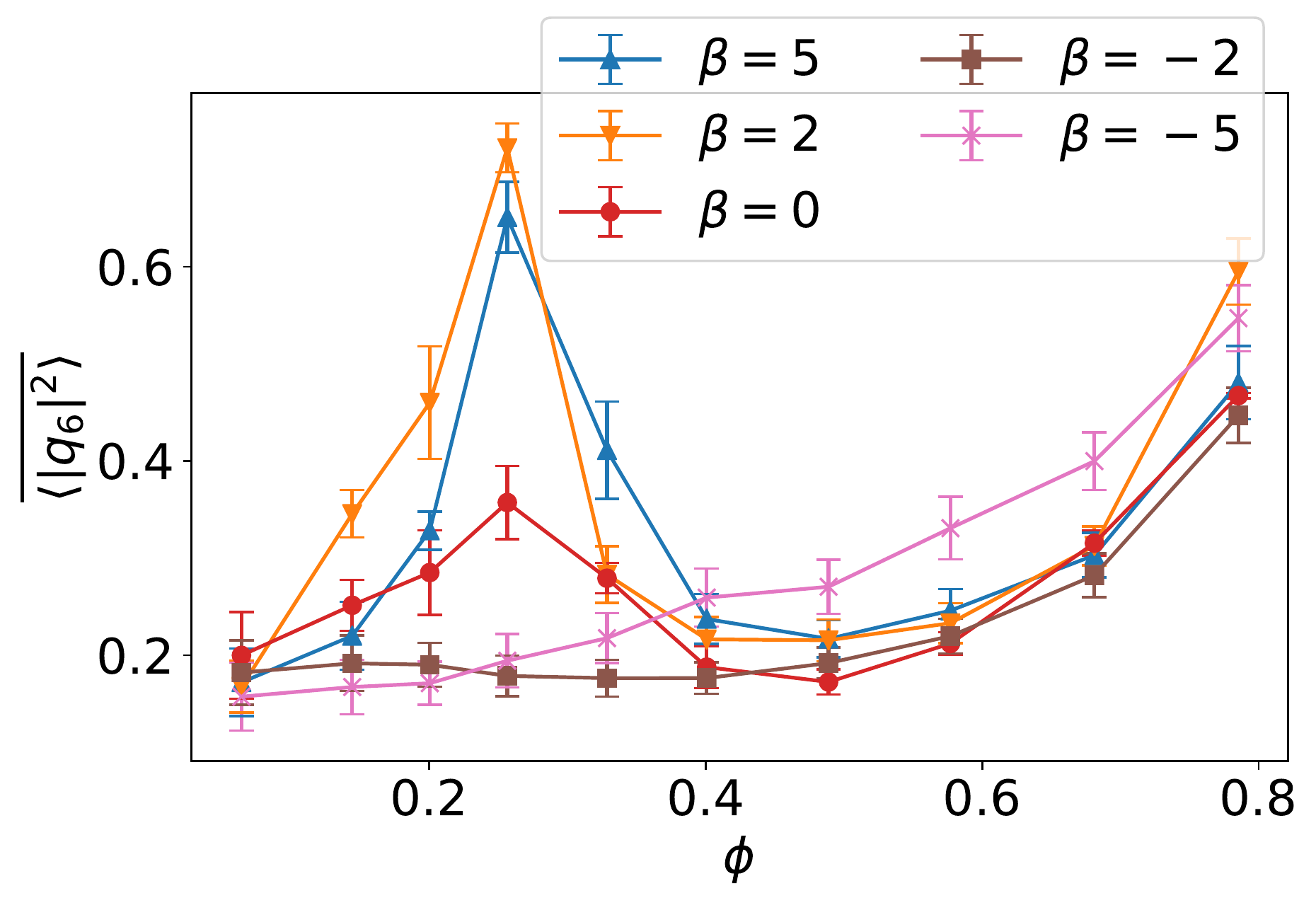}%
\caption{%
The 6-fold bond orientational order parameter
$\overline{\langle\left| q_6^2 \right|\rangle}$ 
plotted versus $\phi$ for different $\beta$ in squirmer monolayers with linear extension $L=112$.
Error bars give the
standard deviation of $\langle\left| q_6\right|^2 \rangle$ 
when averaged over time.
}%
\label{fig:q6_vs_phi}%
\end{figure}%

\subsubsection{Structural order} \label{subsec.structural}
We now present a more quantitative analysis also addressing the missing long-range order in the system. 
To make the ordering visible, we performed Voronoi tessellations for the squirmer monolayers and 
identified pentagons/heptagons as defects in the hexagonal order 
(see snapshot in Fig.~\ref{fig:parameterstudy}, bottom). 
Videos S1 and S9 
in the supplemental material\dag{} illustrate the dynamics of these defects 
for $\beta=0$ and $\beta = 2$ at $\phi = 0.26$.
Clearly, for $\beta = 2$ 
fewer
defects are visible. 
To quantify the local ordering, we introduce the 6-fold bond orientational order parameter 
$\overline{\langle\left| q_6^2 \right|\rangle}$~\cite{Steinhardt:1983, Bialke:2012, Zottl:2014}, 
which averages the local bond order value
\begin{align}
q_6^{(k)}:= 1/|N_k| \sum_{j \in N_k}e^{i6\alpha_{\!k\!j}}
\label{eq.bond_local}
\end{align}
over all squirmers. 
Here, $N_k$ is the number of all Voronoi neighbors of squirmer $k$ 
and $\alpha_{kj}$ is the angle between the horizontal direction 
and the vector connecting squirmers $k$ and $j$.
Note we always use $\langle \ldots \rangle$ to indicate the ensemble average over all squirmers 
and $\overline{\vphantom{a} \ldots \,}$ to indicate the temporal mean. 
In Fig.~\ref{fig:q6_vs_phi} we plot $\overline{\langle\left| q_6^2 \right|\rangle}$ versus $\phi$ for different $\beta$. 
In the region of the hydrodynamic Wigner fluid at $\phi < 0.3$ we see bond orientational order for $\beta > -2$, 
which is most pronounced at $\phi = 0.26$ and $\beta \ge 2$ as already stated. 
The swarming state ($ \beta \le -2$) at small $\phi$ does not show any bond order, 
while a noticeable bond order develops towards the global cluster or swarming state at large $\phi$ for all $\beta$.

\begin{figure}%
\centering%
\includegraphics[width=0.9\linewidth]{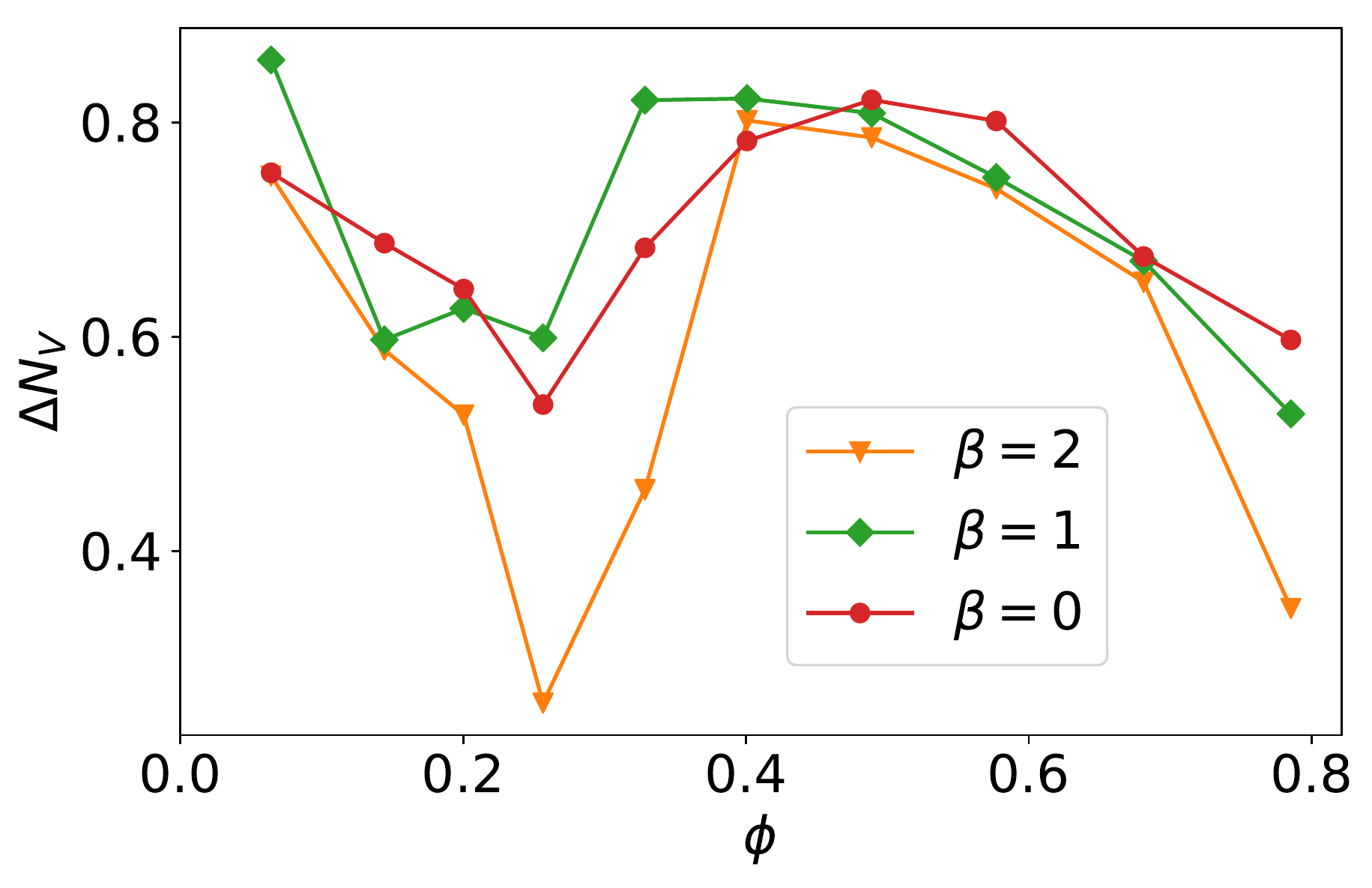}%
\caption{%
The standard deviation from the mean
number of Voronoi neighbors, $\Delta N_V$,
plotted versus $\phi$ for $\beta = 0$, $1$, and $2$.
Linear system size is $L=112$.
}%
\label{fig:std_voro_neighs_vs_phi}%
\end{figure}

To analyze the Voronoi tessellation further, 
we plot in Fig.~\ref{fig:std_voro_neighs_vs_phi} the standard deviation $\Delta N_V$ 
from the mean number number of Voronoi neighbors, 
which is six in all our simulations
as required by the planar surface.
The minimum at $\phi = 0.26$ confirms the earlier observations 
that at this density local hexagonal order is most pronounced.

\begin{figure}%
\centering%
\includegraphics[width=0.9\linewidth]{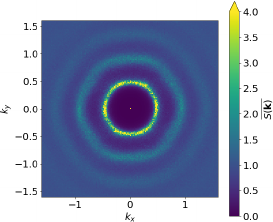}%
\caption{%
Structure factor color-coded in the $k_x, k_y$ plane for $\beta =0$, 
$\phi = 0.26$, and $L= 448$.
}%
\label{fig:structure_factor}%
\end{figure}

Ultimately, to probe the squirmer monolayers for long-range positional order, 
we determine the structure factor
\begin{align}
S(\mathbf{k}) := \frac1N \left\langle\sum_{j,k=1}^N \exp\left[i\mathbf{k} \cdot \left(\mathbf{r}_j - \mathbf{r}_k\right)\right] \right\rangle \; .
\end{align}
In Fig.~\ref{fig:structure_factor} it is color-coded in the $k_x, k_y$ plane for $\beta =0$ at $\phi = 0.26$. 
Clearly, the inner ring at $|\mathbf{k}^{\ast}|$ 
and two weak larger rings indicate that long-range positional order does not exist 
although weak maxima are visible in the inner ring,
which we attribute to the finite size of our simulations.
This justifies the name hydrodynamic Wigner fluid. 
We also checked that a perfectly ordered squirmer monolayer is not stable 
but develops pentagon/heptagon defects in time and ultimately shows the same structure factor. 
For $\beta = 1, 2$
the weak maxima in $S(\mathbf{k})$ are a bit more pronounced and seem to indicate 
the different hexagonal domains visible in the Voronoi tessellation of video S9. 

\begin{figure}%
\centering%
\includegraphics[width=0.9\linewidth]{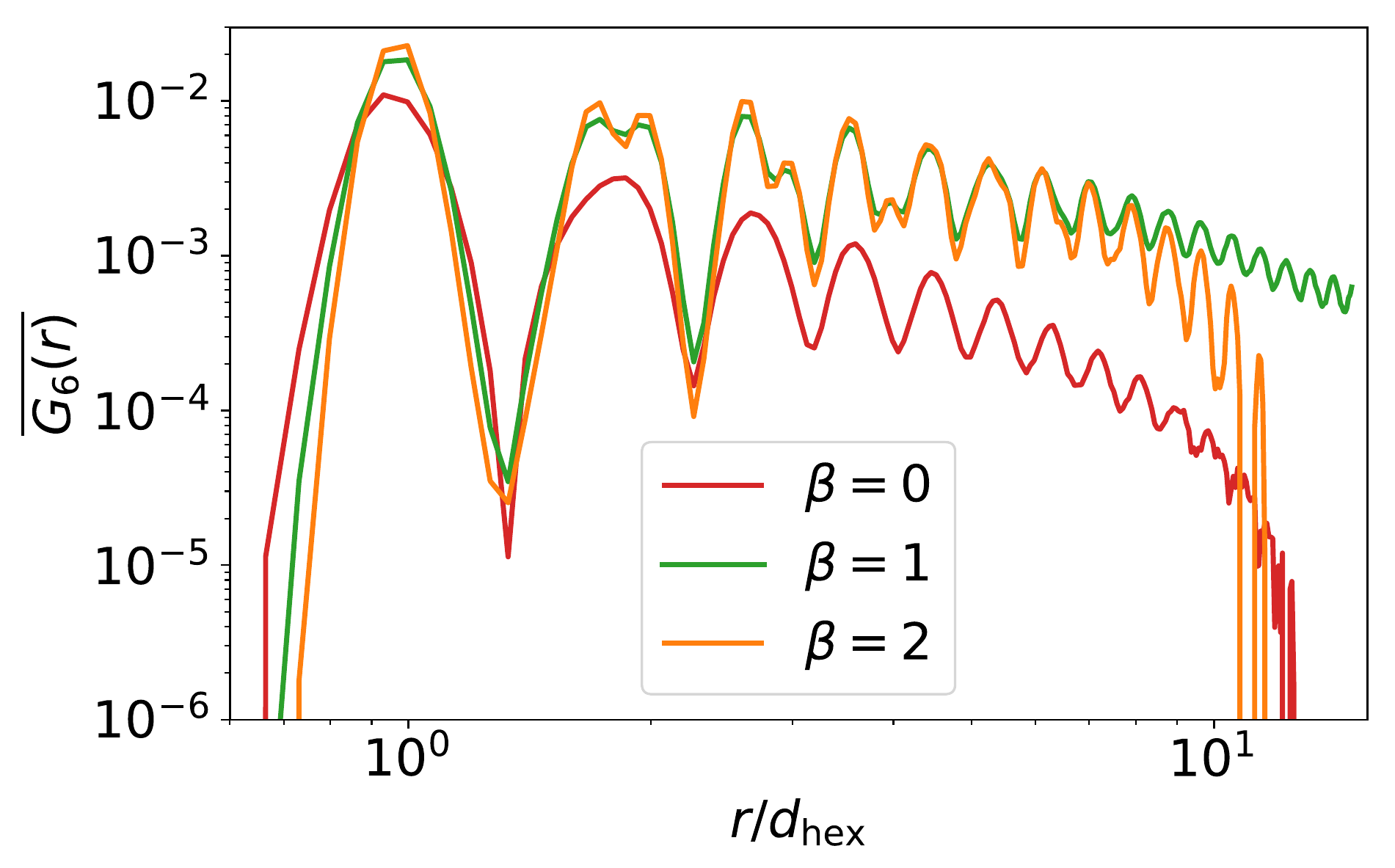}%
\caption{%
Time-averaged 6-fold bond order correlation function plotted versus versus distance $r$ for $\beta = 0, 1$ 
and $2$ at $\phi = 0.26$.
Linear system size is $L= 448$.
}%
\label{fig:loglog_bo_corr_1D}%
\end{figure}

A two-dimensional system allows for a hexatic phase, 
which is fluid but shows long-range correlations in the 6-fold bond order 
as demonstrated in the theory of two-dimensional 
melting~\cite{zahn1999two, strandburg1988two, gasser2010melting, zahn2000dynamic, eisenmann2004anisotropic}. 
The 6-fold bond-order correlation function,
\begin{align}
G_6(\left|\mathbf{r} -\mathbf{r}'\right|) := \langle q_6(\mathbf{r}) q_6(\mathbf{r}') \rangle \, ,
\end{align}
with $q_6(\mathbf{r})$ defined in Eq.~\eqref{eq.bond_local} identifies hexatic order 
by a power-law decay compared to an exponential decay in the liquid state~\cite{gasser2010melting}. 
We plot it in Fig.~\ref{fig:loglog_bo_corr_1D} for $\beta = 0, 1$, and $2$.
Since $\overline{G_6(r)}$ for $\beta = 0, 2$
decays strongly between distances $r/d_\mathrm{hex} = 100$ and $200$,
hexatic order is not present and the hydrodynamic Wigner fluid is in a pure liquid state.
Surprisingly, for $\beta = 1$ the strong decay is not visible. 
The Voronoi tessellation reveals hexagons oriented roughly in the same direction all 
over the simulation box and only small regions with a different mean orientation.
This might cause the observed feature and only simulations with a larger system size
can clarify if it is a finite size effect.
However, at present such simulations are unfeasible also because 
the system is governed by a dynamics slower by at least a factor of ten 
compared to $\beta = 0$, as we will see below.

\subsubsection{Relaxation dynamics}
\begin{figure}%
\centering%
\includegraphics[width=1.0\linewidth]{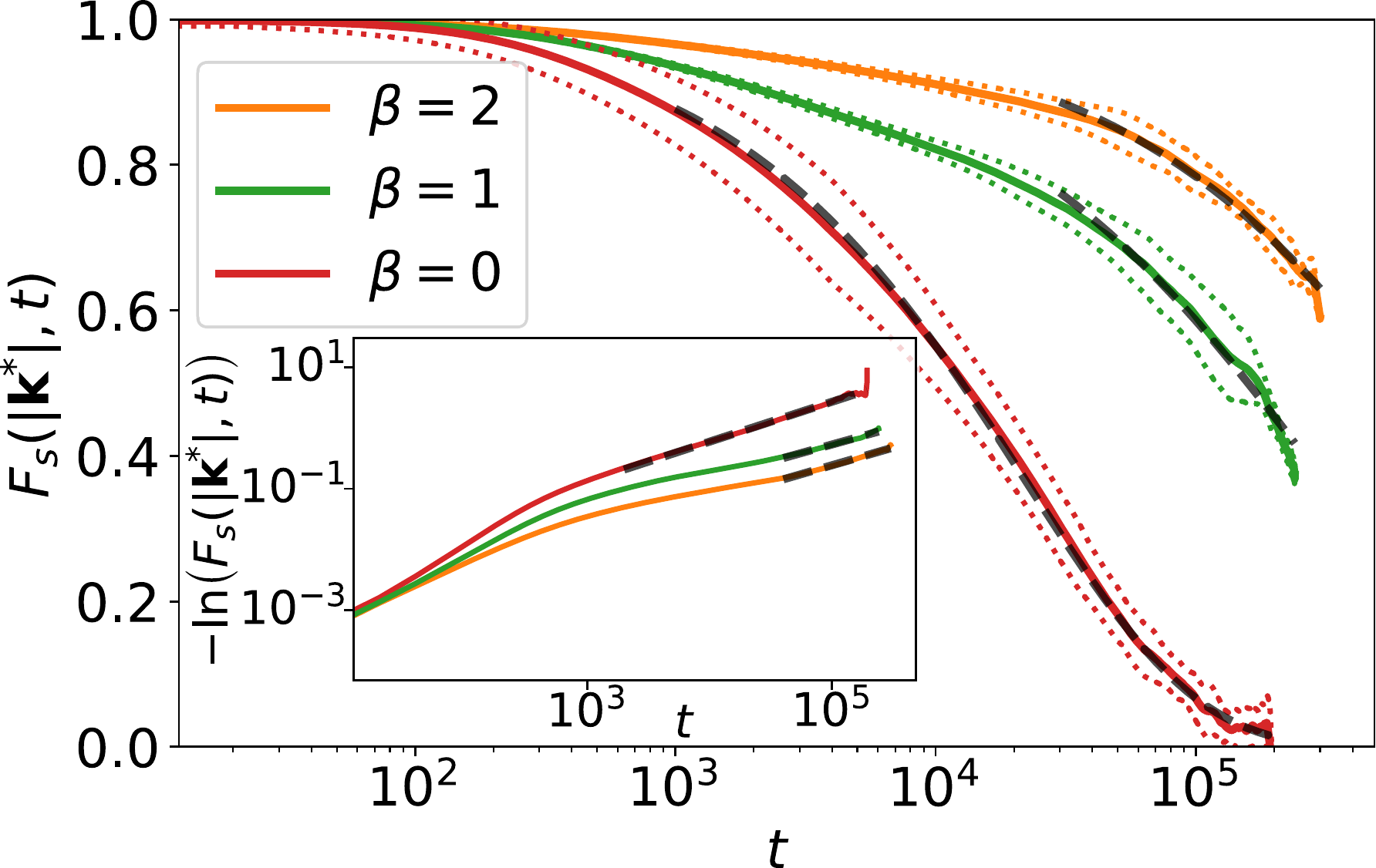}
\caption{%
Self-intermediate scattering function $F_s(|\mathbf{k}^\ast|, t)$
evaluated at the maximum $\mathbf{k}^\ast$ of the structure factor for different $\beta$at $\phi = 0.26$.
Linear system size is $L= 448$.
The black dashed lines are best fits with a stretched exponential.
The doted lines indicate the standard deviation across all data with identical $t$.
Inset: Double-logarithmic plot of $-\ln F_s(| \mathbf{k}^\ast |,t)$.
}%
\label{fig:logx_int_scat_func}%
\end{figure}

Finally, we monitor the relaxation dynamics of the hydrodynamic Wigner fluid 
by looking at the self-intermediate scattering function
\begin{align}
F_s(\mathbf{k}, t) := \frac1N \left\langle\sum_{j=1}^N \exp\left[i\mathbf{k} \cdot \left(\mathbf{r}_j(t) - \mathbf{r}_j(0)\right)\right] \right\rangle \; ,
\end{align}
which probes the motional dynamics of squirmers on lengths associated with the wave vector $\mathbf{k}$. 
Due to the overall rotational symmetry as illustrated by the structure factor, 
the self-intermediate scattering function only depends on the magnitude $\left|\mathbf{k}\right|$. 
Therefore, we average over all wave vectors with the same $\left|\mathbf{k}\right|$ 
when determining $F_s(\left| \mathbf{k} \right|\!, t)$.
In Fig.~\ref{fig:logx_int_scat_func} we plot $F_s(\left|\mathbf{k}^\ast\right|\!, t)$ for three different $\beta$
using the wave number $\left|\mathbf{k}^\ast\right| = 2\pi / d$ 
where the structure factor is maximal.
Thus, $d \approx d_{\mathrm{hex}}$ is the distance between nearest-neighbor squirmers. 
We observe relaxational dynamics that slows down with increasing $\beta$.
It demonstrates that on lengths comparable to $d$ 
the motion of single squirmers becomes decorrelated. 
Due to constraints in the possible simulation time, 
the relaxation for $\beta=1, 2$ is not complete. 

Interestingly, after some initial decay the self-intermediate 
scattering function, especially for $\beta=0$, 
can be fitted by a stretched exponential,
\begin{align}
f(t) = e^{-(t/\tau)^\alpha} \; .
\end{align}
Typically, this indicates a more complex relaxation process. 
For $\beta = 0$ we find the exponent $\alpha = 0.66$ as indicated in the inset of Fig.~\ref{fig:logx_int_scat_func}. 
Stretched exponentials are observed in the $\alpha$-relaxation of (colloidal) glasses close to the glass 
transition~\cite{Sciortino_2005, Pusey_2008, brambilla2009probing, Hunter_2012, berthier2011theoretical}. 
It is preceded by the
$\beta$-relaxation\footnote[5]{Note that this has nothing to do with the squirmer parameter $\beta$.}%
, which enters a plateau before $\alpha$-relaxation sets in. 
We do not observe such a plateau.
However, already at $\beta = 0$ we can identify a different initial relaxation process in $F_s(\left| \mathbf{k} \right|\!, t)$,
which cannot be fit by the stretched exponential as the inset demonstrates.
This process is even better visible for $\beta = 1$ and $2$.

\begin{figure}%
\centering%
\includegraphics[width=1.0\linewidth]{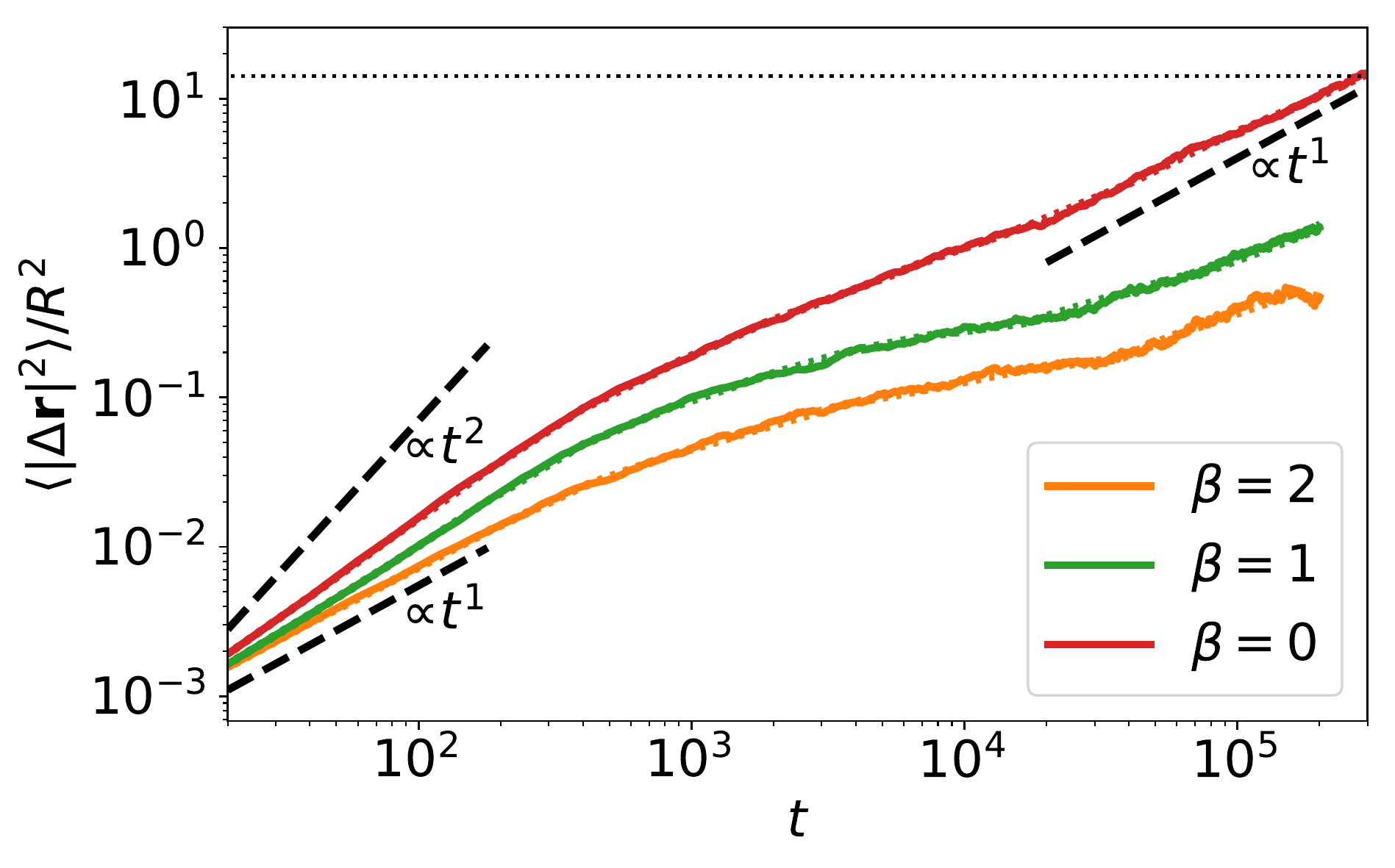}%
\caption{%
Mean squared displacement (in units of $R^2$)
of neutral and puller squirmers ($\beta = 1$ and 2)
at density $\phi = 0.26$.
Linear system size is $L= 448$.
The horizontal dashed line belongs to the distance $d_\mathrm{hex}/R = 3.75$.
Note we find that the distribution of squared displacements, $P(\Delta \mathbf{r}^2)$, 
for $\beta = 0$ closely matches an exponential distribution for all $t$, where
the standard deviation is equal to the mean value, 
$\langle \Delta \mathbf{r}^2 \rangle = \sigma \left(\Delta \mathbf{r}^2 \right)$.
Thus, in our system with $N=1024$, the standard error of the mean
is
$\sigma \left(\Delta \mathbf{r}^2 \right)/\sqrt{N} = \langle \Delta \mathbf{r}^2 \rangle / 32$.
We plot the standard error as dotted lines around $\langle \Delta \mathbf{r}^2 \rangle$ also for $\beta = 1,2$. 
They are hardly distinguishable from the main curves.
}%
\label{fig:msd_vs_time_i3}%
\end{figure}

Close to the colloidal glass transition $\alpha$-relaxation is associated 
with a colloid leaving the cage formed by surrounding colloids. 
We anticipate a similar dynamics as illustrated in Fig.~\ref{fig:msd_vs_time_i3}, 
where we plot the mean squared displacement versus time.
When calculating $\langle\left|\Delta \mathbf{r}\right|^2 \rangle$ we first subtract any global
drift motion of all squirmers.
For $\beta = 0$ squirmers move beyond the squirmer-squirmer distance $d_\mathrm{hex}$ 
leaving the cage formed by neighboring squirmers. 
For $\beta =1$ and $2$ this process is not completed in the available simulation time
in agreement with Fig.~\ref{fig:logx_int_scat_func}. 
Interestingly, the mean-square displacement is (super)diffusive
for small times, becomes subdiffusive at intermediate times, 
when the squirmer starts to feel its neighbors,
and for large times for $\beta = 0$ approaches diffusive dynamics.
Thus, squirmers in a hydrodynamic Wigner fluid perform glassy dynamics 
reminiscent of what is observed close to  the glass transition.

\subsection{Clusters\label{sec:clusters}}
We now discuss in more detail aspects of the state of fluctuating chains and clusters, 
the kissing and the global cluster states, 
which we introduced in the state diagram of Fig.~\ref{fig:parameterstudy}. 
In Sect.~\ref{sec:hexorder} we noted that starting with the hydrodynamic Wigner fluid and 
increasing density beyond $\phi = 0.26$ squirmers touch each other 
and form pairs due to an effective short-range hydrodynamic attraction.
Further squirmers can join to form chains and clusters but this process is reversible. 
In the state of fluctuating clusters 
squirmers detach due to stochastic fluctuations and the cluster breaks apart, 
while in the kissing state the breakup looks more deterministic.

\subsubsection{Cluster size: Mean values and distributions}
The formation of clusters leaves more space to surrounding squirmers and 
reduces their tendency to attach to other squirmers. 
This mechanism leads to a steady state, 
which we characterize by the mean value and the distribution of the number $N_{cl}$ of squirmers in a cluster. 

\begin{figure}%
\centering%
\includegraphics[width=1.0\linewidth]{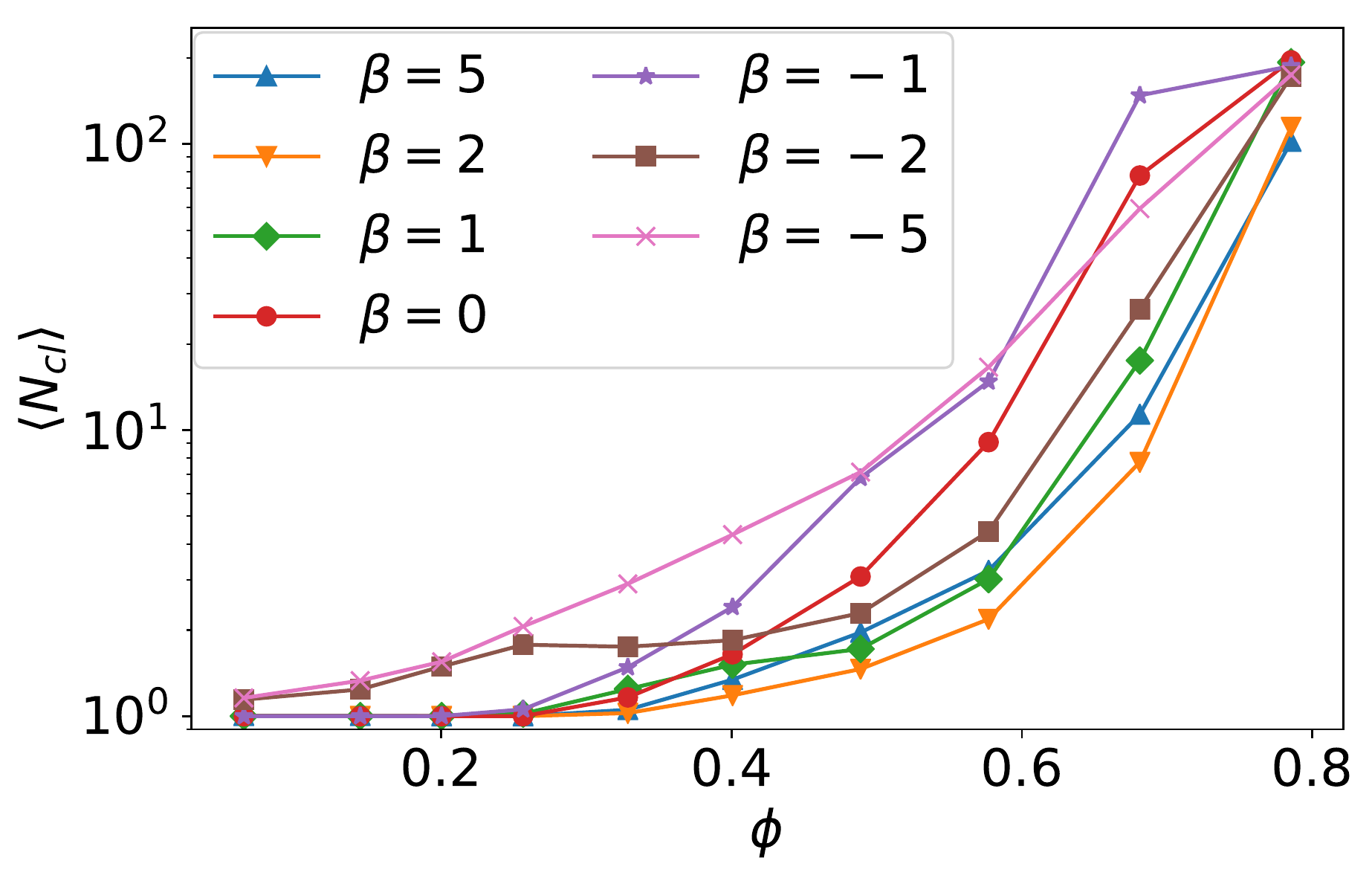}%
\caption{%
The mean cluster size $\langle N_{cl} \rangle$ 
plotted versus density $\phi$ for different squirmer types $\beta$.
}%
\label{fig:mean_cluster_size}%
\end{figure}

The formation of clusters, their sizes and stability depend on 
the hydrodynamic interactions between the squirmers.
For all $\beta$ the probability of squirmers to touch each other and 
thereby form clusters grows with area density $\phi$, 
as the mean cluster size $\langle N_{cl} \rangle$, 
plotted in Fig.~\ref{fig:mean_cluster_size}, demonstrates.
To acquire the statistical data for calculating mean values and also cluster-size distributions, 
we define squirmers to be in the same cluster if the gap between them is smaller than $0.1R$, 
\textit{i.e.}, the distance of their centers is below $2.1R$.
For all squirmer types $\beta$ the mean cluster size ultimately grows faster than exponentially with increasing $\phi$.
Of course, in the hydrodynamic Wigner fluid ($\beta > -2$) the mean cluster size is one and 
then increases beyond $\phi = 0.26$ when the states of kissing or fluctuating clusters are entered. 
In the swarming state of strong pushers (see $\beta = - 5$) and 
in the fluctuating cluster state of weaker pushers 
(see $\beta = -2$) the mean cluster size is larger than 1 even for small $\phi$. 
Here, the squirmers are tilted against the wall normal so that 
they constantly move along the bottom wall. 
In this dynamic environment they bump into each other and form transient clusters.

\begin{figure}%
\centering%
\includegraphics[width=1\linewidth]{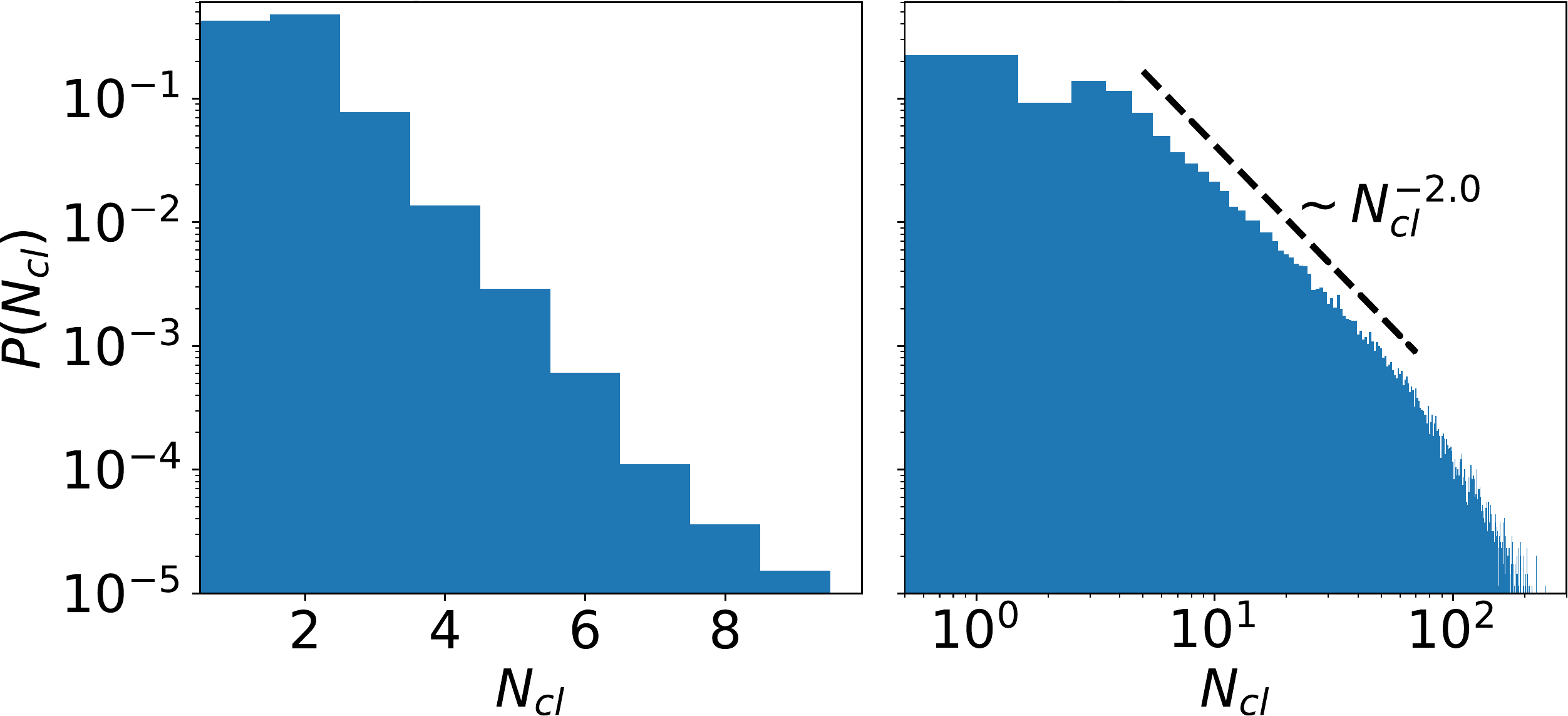}%
\caption{%
Distributions of cluster sizes, 
$P(N_{cl})$, of neutral squirmers ($\beta = 0$).
Linear system size is $L = 448$.
\textbf{left}: $\phi=0.4$, note the semi-logarithmic plot;
\textbf{right}: $\phi=0.58$, note the double-logarithmic plot.
}%
\label{fig:cluster_dists}%
\end{figure}

We further analyzed the distribution $P(N_{cl})$ of cluster sizes for neutral squirmers and found several
characteristic shapes (see Fig.~\ref{fig:cluster_dists} for two of them).
Of course, for $\phi \leq 0.26$ all clusters have size one (distribution not shown). 
Above $\phi = 0.26$ in the fluctuating cluster state 
the cluster-size distribution has exponential form (see Fig.~\ref{fig:cluster_dists}, left), 
where the mean size grows with $\phi$ as discussed before.
This shape persists until the largest cluster of squirmers is close to the percolation transition, 
where the cluster spans the whole system. 
Below but close to this transition (see Fig.~\ref{fig:cluster_dists}, right) 
the cluster-size distribution takes the form of a power law with an exponential cutoff. 
The exponent of the power law $P(N_{cl}) \sim N_{cl}^{-\tau}$ has the value $\tau \approx 2.0$, 
which nicely agrees with the theoretical value $187/91 \approx 2.05$ 
of two-dimensional percolation~\cite{stauffer1979scaling, stauffer2014introduction}.
Finally, at high $\phi$ the distribution $P(N_{cl})$ is bimodal (not shown).
It results from the dominant percolating cluster of fluctuating size, 
which scales with $L^2$ at constant $\phi$, 
and a distribution of smaller clusters, 
which break away from and merge with the dominant cluster dynamically.

\subsubsection{Kinetics of fluctuating pairs/trimers, and kissing\label{subsec:pairs_trimers_kissing}}
Figure~\ref{fig:parameterstudy}, bottom shows an example of a system 
in the fluctuating cluster state with abundant chains and 
a trimer, where three squirmers form a nearly equilateral triangle. 
Also in the kissing state such motives are found. 
We concentrate here on pairs or trimers and characterize their kinetics further.

\begin{figure}%
\centering%
\includegraphics[width=1.0\linewidth]{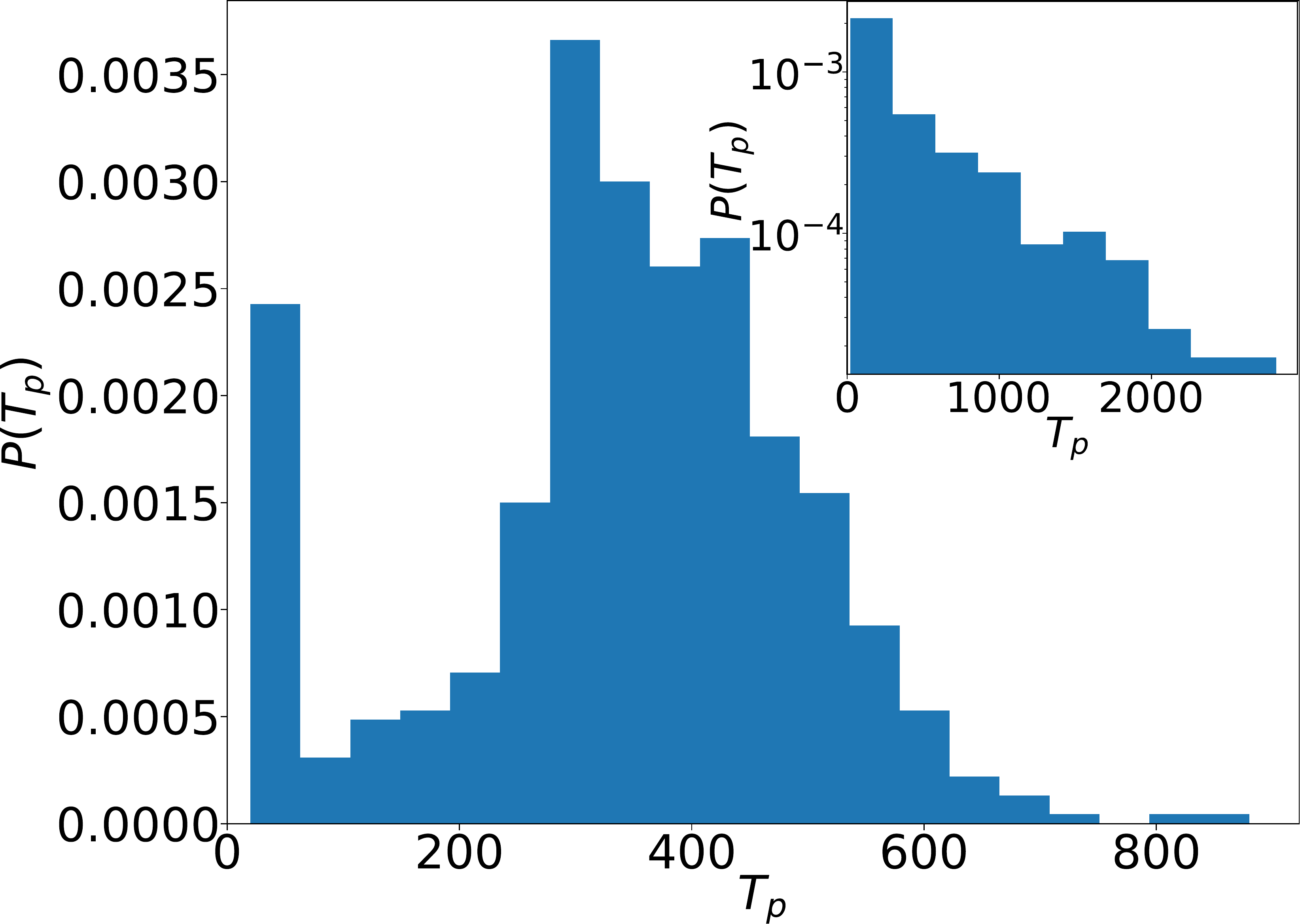}
\caption{%
The distribution of pair lifetimes, $P(T_p)$,
for puller squirmers ($\beta = 2$) at $\phi = 0.33$ in the kissing state.
Linear system size is $L = 112$.
Inset: Pair lifetime distribution in the fluctuating cluster state for $\beta = -2$ at $\phi = 0.33$, note the semi-logarithmic plot.
}%
\label{fig:pair_lifetimes}%
\end{figure}

In the fluctuating cluster state pairs form when the orientations of the squirmers are tilted towards each other
so that they swim against each other (see video S10 in the supplemental material\dag{}).
The pairs break up again 
when nearby squirmers approach
or when orientational fluctuations tilt the orientations away.
The events of a breakup occur stochastically and seem to be independent from each other.
Then, we expect them to follow a Poissonian process like radioactive decay.
Indeed, the distribution of pair lifetimes is roughly exponential 
as the inset of Fig.~\ref{fig:pair_lifetimes} demonstrates.

In the kissing state the breakup of pairs and trimers follows a different process. 
This is obvious from the distribution of pair lifetimes 
plotted in Figure~\ref{fig:pair_lifetimes} for puller squirmers ($\beta = 2$) at $\phi = 0.33$.
The distribution has a clear maximum at a non-zero lifetime $T_p$.
Indeed, video S2 in the supplemental material\dag{} 
and the graphic representation in Fig.~\ref{fig:parameterstudy}, 
bottom for the formation and breakup of a pair hints to a mostly deterministic process. 
As squirmers approach, their orientations are tilted towards each other.
However, this configuration is unstable. 
As soon as they touch, they turn to the side, 
pass each other, and thereby separate to find another nearby puller. 
Due to this scenario, we called the dynamical state ``kissing''.
Interestingly, similar behavior occurs for trimers
(marked with a red circle in video S2 in the supplemental material\dag{}).

Finally, we note that for the large density of $\phi = 0.68$ and $\beta = -1$ 
we even observe symmetric clusters with seven squirmers 
as part of larger groupings
(see,
\textit{e.g.}, bottom left in the beginning of 
video S11 in the supplemental material\dag{}). 
They spontaneously switch between rotating and non-rotating states.

\subsection{Swarming\label{sec:swarming}}
\begin{figure}%
\centering%
\includegraphics[width=1.0\linewidth]{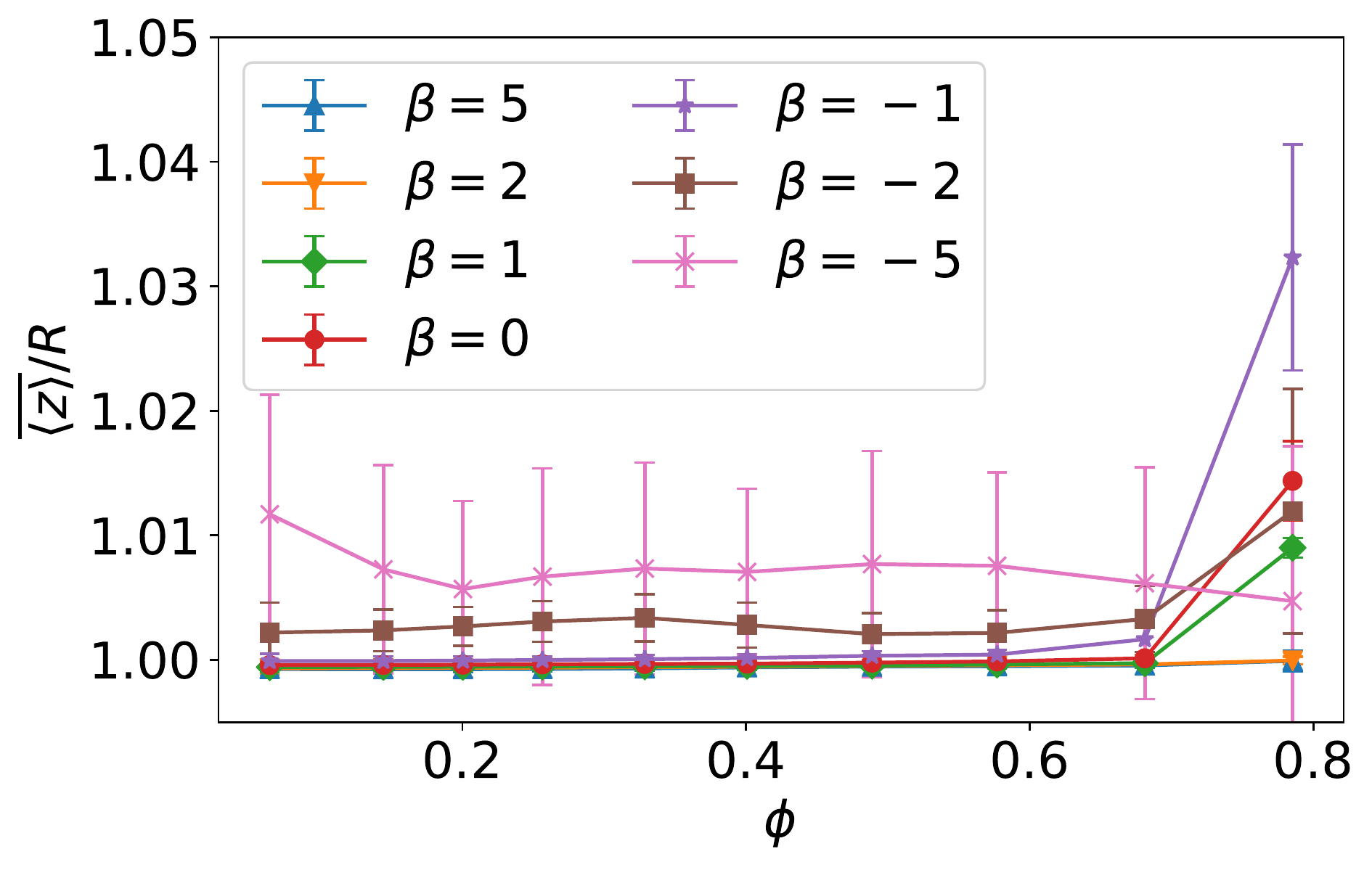}%
\caption{%
Mean height of squirmers above the bottom wall in units of squirmer radius $R$
plotted versus density $\phi$ for different squirmer types $\beta$.
Error bars indicate the temporal standard deviation of the recorded heights.
}%
\label{fig:dist_wall}%
\end{figure}

\begin{figure}%
\centering%
\includegraphics[width=1.0\linewidth]{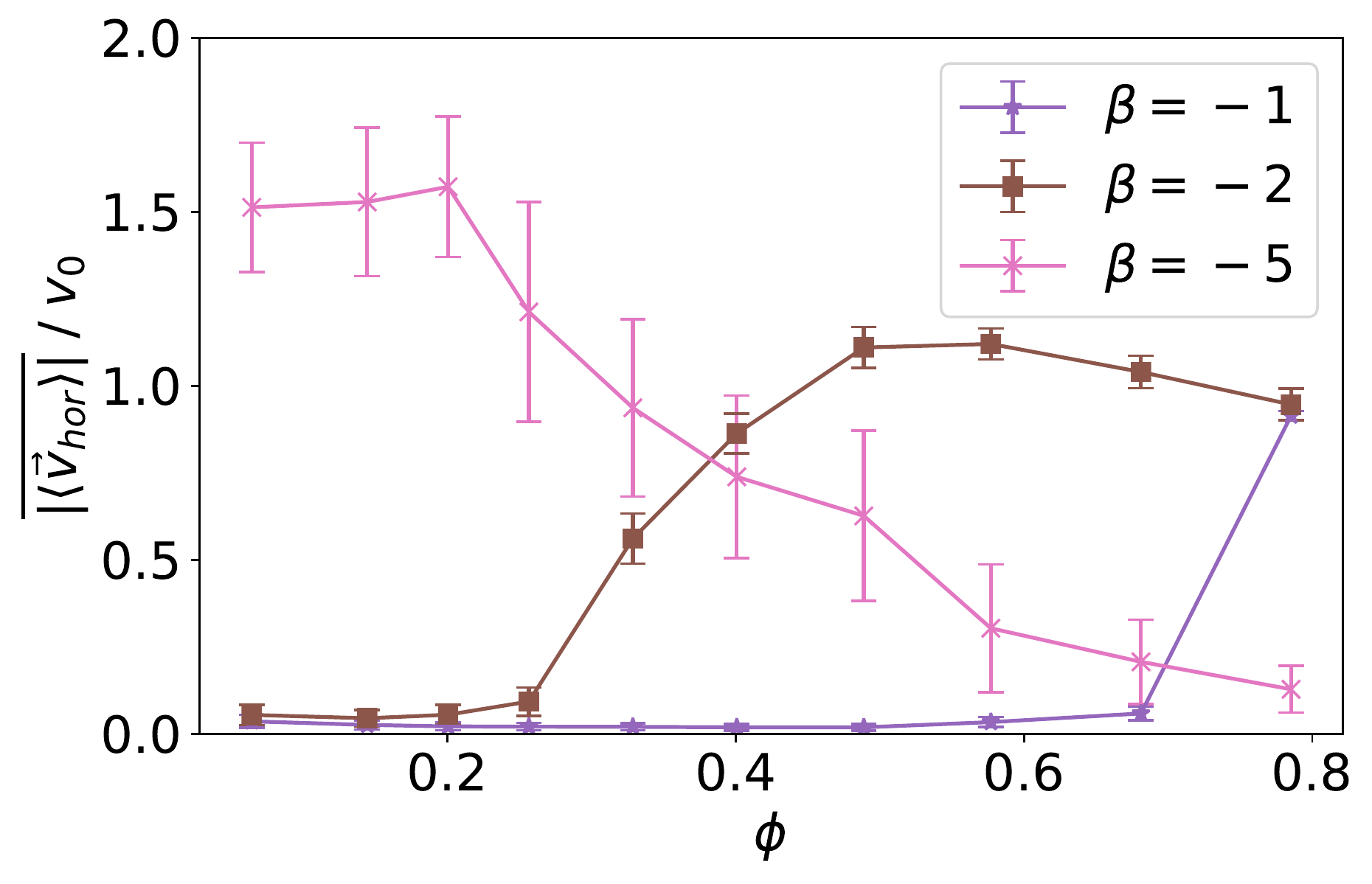}%
\caption{%
Mean collective horizontal speed plotted versus density $\phi$ for squirmer type $\beta = -1$, $-2$, and $-5$.
Error bars indicate the temporal standard deviation of the recorded horizontal velocities.
}%
\label{fig:swarming}%
\end{figure}

For $\beta \le -2$ but also for $\beta = -1$ at large densities
the pusher orientation tilts against the normal so that it moves along the bottom wall. 
This observation correlates very nicely with the mean height of the squirmer above the wall, 
which is plotted in Fig.~\ref{fig:dist_wall} versus density $\phi$ for different $\beta$.
While for $\beta > -2$ squirmers sit on the surface, 
for $\beta \le -2$ the flow field due to the tilted orientation lifts them up by a small amount. 
We note that the tilted orientation fits to the expectation 
that for pushers there has to be a transitional region between the upright orientation at the wall 
as calculated in lubrication theory and the parallel far-field orientation~\cite{lintuvuori2016hydrodynamic, Ruhle:2018}.
For small densities and squirmer type $\beta$ around $-2$ 
we observe random motion of the squirmers in the fluctuating chain state
as indicated in the state diagram of Fig.~\ref{fig:parameterstudy}. 
For stronger pushers the in-plane velocities align and the squirmers show swarming 
(see video S5 in the supplemental material\dag{}).
We note that in this state the in-plane velocities point in the same direction as the 
in-plane orientations of the squirmers (data not shown). As we will see in Fig.~\ref{fig:swarming}
strong pushers ($\beta = -5$) show a mean in-plane velocity of $1.5 v_0$ at low densities, which cannot be explained
by the mean tilt of a single squirmer against the wall normal (see Fig.~\ref{fig:single_sq_orientation}). So the 
enhanced swarming velocity is also a collective effect.

Figure~\ref{fig:swarming} quantifies the swarming by plotting the mean collective horizontal speed 
$\overline{\left|\langle\mathbf{v}_\text{hor}\rangle\right|}$, 
where $\mathbf{v}_\text{hor} := (v_x, v_y)$. 
Here it is important that first the ensemble average of the in-plane velocity vector is taken before
the absolute value in order to identify a collective swarming direction.
Thus, swarming or collective motion in a preferred direction is measured.
We consider the system to display swarming if 
$\overline{\left|\langle\mathbf{v}_\text{hor}\rangle\right|}/v_0 > 0.5$.
The mean collective horizontal speed reflects the different states in the state diagram of Fig.~\ref{fig:parameterstudy}. 
Since the slope of the curves in Fig.~\ref{fig:swarming} are rather steep around 
$\overline{\left|\langle\mathbf{v}_\text{hor}\rangle\right|}/v_0 = 0.5$, changing the criterion will not strongly affect 
the schematic boundaries in the state diagram.
For $\beta = -1$ it is zero, which corresponds to the Wigner fluid and the adjacent fluctuating cluster state.
Only at the largest densities a noticeable collective horizontal speed indicates swarming.
For $\beta = -2$ the in-plane velocities have random orientation at small densities and then for $\phi \ge 0.3$ align 
with each other in the swarming state. 
Interestingly, the in-plane velocity of strong pushers decreases with increasing $\phi$ above $\phi=0.2$.
Above $\phi \approx 0.58$ a global direction in the collective horizontal velocity does no longer exist and 
the squirmers perform what we call 
chaotic swarming (see video S6 in the supplemental material\dag{}). 
The flow field of strong pushers possesses vortices close to the squirmer surface~\cite{evans2011orientational}, 
which rotate nearby pushers and thereby more and more destroy the alignment of the in-plane velocity for increasing $\phi$. This randomization is also evident from the large error bars in Fig.~\ref{fig:swarming}, 
which indicate the strength of temporal fluctuations of $\left|\langle\mathbf{v}_\text{hor}\rangle\right|$. 
Finally, we note that in the vicinity of the wall swarming pushers can move faster than pushers in a bulk fluid.

\section{Conclusions\label{sec:discussion}}
Using hydrodynamic simulations with the MPCD method, 
we studied a monolayer of squirmers that 
forms under strong gravity at the bottom surface of a container. 
The squirmers interact with each other via their self-generated flow fields and 
thereby induce several dynamic states, 
which we identified by varying squirmer density and squirmer type.
The most interesting state is certainly the hydrodynamic Wigner fluid that 
neutral squirmers, pullers, and weak pushers form at low to medium densities. 
The squirmers have an upright orientation and push their neighbors away through their flow fields.
They thereby create a structure with local hexagonal order, 
which nevertheless is fluid and does not show long-range hexatic order. 
However, we identified a non-trivial relaxation of the self-intermediate scattering function 
that follows a stretched exponential. 
This is reminiscent of what is seen in the $\alpha$-relaxation of a glass-forming system 
close to the glass transition. 
For example, a Wigner glass in a colloidal system with strong electrostatic repulsion exists~\cite{Bonn99}. 
However, in our case we cannot just increase the strength of the hydrodynamic repulsion 
by increasing density since then the squirmers start to hydrodynamically attract each other.

If density is increased starting from the Wigner fluid, 
squirmers enter the state of fluctuating clusters or, 
for medium to strong pullers, the kissing state. 
Fluctuating clusters also exist for medium pushers at small to medium densities. 
Both states show an exponential cluster size distribution 
but they differ in the kinetics how they break up. 
Fluctuating pairs break up by stochastic events due to orientational fluctuations or 
due to approaching nearby squirmers, which roughly gives an exponential lifetime distribution. 
In contrast, the distribution in the kissing state is peaked at a finite time, 
which hints to a more deterministic process. 
At even larger densities the cluster size distribution becomes 
algebraic close to the percolation transition 
and ultimately reaches a bimodal shape in the global cluster state where
most squirmers are part of a percolating cluster. 
Finally, the orientation of strong pushers and also weaker pusher at large densities 
tilts against the wall normal. 
The in-plane velocities align and thereby form the swarming state, 
which for very strong pushers at large densities becomes chaotic.

In a recent work hydrodynamically interacting squirmers experience an aligning torque towards a bounding wall~\cite{shen2019hydrodynamic}. Similar to our results, the authors detect a variety of different states.
So, it would be interesting to identify an experimental system to study structure formation close 
to a bounding wall under defined conditions, which in our case would mean strong gravity. Possible realizations of squirmers
in such a system are Volvox algae and active emulsions~\cite{maass2016swimming}.\\

\begin{acknowledgments}
We thank Peter Keim for insightful discussion.
This project was funded by Deutsche Forschungsgemeinschaft through the research training group GRK 1558
and priority program SPP 1726 (grant number STA352/11). 
The authors acknowledge the North-German Supercomputing Alliance (HLRN) for providing HPC resources that have contributed to the research results reported in this paper.
\end{acknowledgments}

\bibliographystyle{apsrev4-1}
\bibliography{bibliography}

\end{document}